\documentclass[notitlepage,aps,prl,twocolumn,superscriptaddress,amsmath,showpacs,tightenlines,longbibliography,10pt,footnote]{revtex4-1}

\usepackage{ulem}
\usepackage{bm}
\usepackage{amssymb}
\usepackage{amsmath}
\usepackage{dcolumn}
\usepackage{graphicx}
\usepackage{mathrsfs}
\usepackage{subfigure}
\usepackage{booktabs}
\usepackage{color}
\usepackage{docmute}
\usepackage{hyphenat}
\usepackage{CJKutf8}
\usepackage{multirow}
\usepackage{array}
\usepackage{threeparttable}
\usepackage[mathscr]{euscript}

\usepackage{graphicx}
\usepackage{subfigure}
\usepackage{tikz}
\usetikzlibrary{positioning}
\usepackage{diagbox}

\newcommand{\bea}{\begin{eqnarray}}
\newcommand{\eea}{\end{eqnarray}}
\newcommand{\beq}{\begin{equation}}
\newcommand{\eeq}{\end{equation}}
\newcommand{\nn}{\nonumber}
\def\/{\over}

\usepackage{url}
\usepackage[colorlinks]{hyperref}
\hypersetup{%
	plainpages=true,
	breaklinks=true,
	hypertexnames=false,
	pageanchor=true,
	bookmarksnumbered=true,
	colorlinks=true,
	linkcolor={blue},
	citecolor={red},
	urlcolor={blue},
	anchorcolor={black}
}

\begin{document}

\title{Molecular entanglement as a signature of the Unruh effect}
\author{Yuebing Zhou}
\thanks{These authors contributed equally to this work.}
\affiliation{Department of Physics, Key Laboratory of Low Dimensional Quantum Structures and Quantum Control of\\ Ministry of Education, and Hunan Research Center of the Basic Discipline for Quantum Effects and Quantum Technologies, Hunan Normal University, Changsha, Hunan 410081, China}
\affiliation{Department of Physics, Huaihua University, Huaihua, Hunan 418008, China}
\author{Jiawei Hu}
\thanks{These authors contributed equally to this work.}
\affiliation{Department of Physics, Key Laboratory of Low Dimensional Quantum Structures and Quantum Control of\\ Ministry of Education, and Hunan Research Center of the Basic Discipline for Quantum Effects and Quantum Technologies, Hunan Normal University, Changsha, Hunan 410081, China}\author{Hongwei Yu}
\email[Corresponding author: ]{hwyu@hunnu.edu.cn}
\affiliation{Department of Physics, Key Laboratory of Low Dimensional Quantum Structures and Quantum Control of\\ Ministry of Education, and Hunan Research Center of the Basic Discipline for Quantum Effects and Quantum Technologies, Hunan Normal University, Changsha, Hunan 410081, China}

\begin{abstract}

The Unruh effect predicts that a uniformly accelerated observer perceives the vacuum seen by an inertial observer as a thermal bath at a temperature proportional to its proper acceleration. This phenomenon is often regarded as a flat spacetime ``cousin" of Hawking radiation. In this Letter, we first study the entanglement dynamics of a quantum system composed of two polarizable two-level subsystems undergoing centripetal acceleration in a vacuum. We demonstrate that the system's steady state can be entangled irrespective of the initial state, a distinct characteristic attributable to the circular manifestation of the Unruh effect. Through meticulous analysis, we then propose that this phenomenon can feasibly be detected using state-of-the-art optomechanical technologies, particularly with a quantum system of two molecules.

\end{abstract}

\maketitle

\emph{Introduction}. A quantum system in its ground state, when subjected to uniform acceleration in a vacuum, may transition to excited states similarly to how it would behave in a thermal bath at a temperature proportional to its proper acceleration. 
 This  phenomenon is known as the Unruh effect \cite{Fulling1973,Davies1975,W. G. Unruh}.  
Drawing from the equivalence principle concerning acceleration and gravity, the Unruh effect is often considered  as a flat spacetime ``cousin" of the Hawking radiation \cite{Hawking74,Hawking75}, which plays a crucial role in the comprehensive understanding of quantum gravity. Consequently, 
 direct experimental confirmation of the Unruh effect is viewed as a significant scientific goal of our time.
 
 Numerous theoretical proposals for detecting the Unruh effect have emerged \cite{Pisin99,Habs06,Scully03,Martin11,Hu12},  including approaches such as employing ultraintense lasers to achieve substantial acceleration \cite{Pisin99,Habs06},  using high-Q cavities to enhance the ratio of excitation to de-excitation rates in atoms \cite{Scully03}, and taking Berry's phase as an alternative physical observable for detection \cite{Martin11,Hu12}. 
 
In addition to uniform acceleration scenarios, it has been shown that an observer undergoing circular acceleration would also perceive radiation, albeit with a spectrum that deviates from the Planckian distribution \cite{Bell83,Bell87,Unruh98}. 
This variation is referred to as the circular Unruh effect, and several theoretical frameworks have been proposed to observe this effect \cite{Bell83,Bell87,Unruh98,Rogers88,Rad2012,Lochan20,Arya2022,Arya2023}. 

Moreover,  attempts to simulate the Unruh effect have been carried out in both linear~\cite{Retzker2008} and circular~\cite{Gooding2020} settings using analog gravity platforms such as Bose-Einstein condensates. In these systems, the speed of sound replaces the speed of light as the characteristic propagation speed of perturbations, allowing relativisticlike quantum field effects to be emulated in a controllable, nonrelativistic environment. Despite these efforts, a conclusive experimental verification of the Unruh effect remains outstanding.

\begin{figure}[!htbp]
\centering
\includegraphics[width=0.3\textwidth]{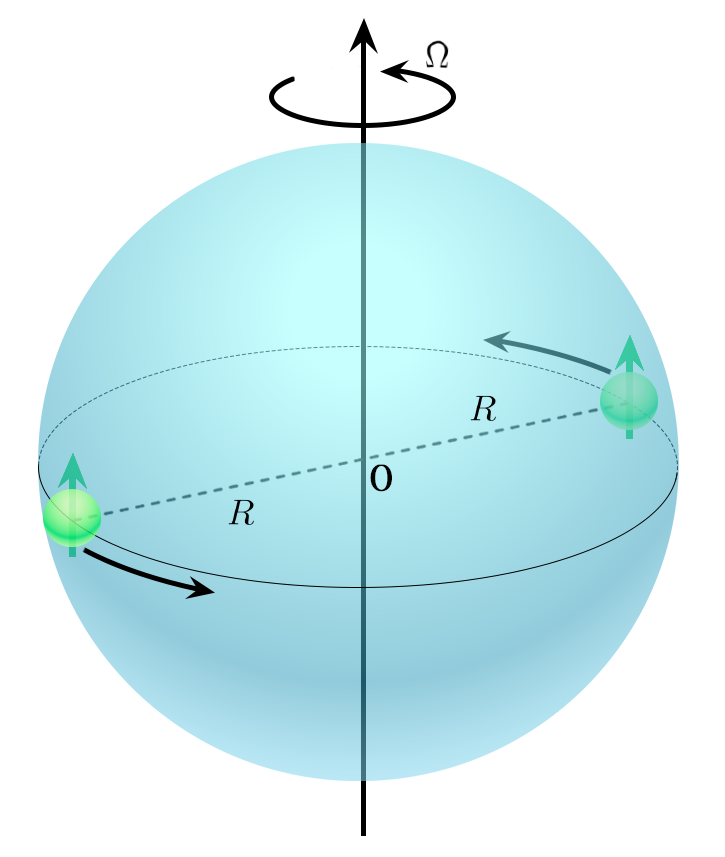}
\caption{\label{model} 
Two two-level quantum systems are attached to the two ends of a diameter of an optically levitated hyperfast rotating nanosphere with radius $R$ and angular frequency $\Omega$. }
\end{figure}

In this Letter, we propose detecting the genuine  circular Unruh effect, rather than an analog version, by measuring the steady-state entanglement of a pair of molecules attached to a hyperfast rotating nanoparticle. First, we show in general that for a quantum system consisting of two neutral but polarizable two-level subsystems under centripetal acceleration, the steady state is entangled when the rotation angular frequency and the orbital radius are properly chosen. Notably, this entanglement can persist in the steady state, regardless of the initial state. This indicates that a ground-state quantum system in vacuum can be excited when it undergoes centripetal acceleration, as otherwise, the two two-level subsystems would remain in their ground states and could not become entangled. Therefore, the steady-state entanglement can be regarded as evidence of the circular Unruh effect, and we propose detecting the circular Unruh effect using this steady-state entanglement.   

While the steady state of a single rotating two-level system contains information about the circular Unruh effect, we instead use steady-state entanglement as a probe, as it represents a distinctive quantum signature unique to the circular Unruh effect. Crucially, such steady-state entanglement arises only when the detected radiation spectrum deviates from a Planckian distribution \cite{Zhou2022}, as is the case for circular acceleration \cite{Bell83,Bell87,Unruh98}. In contrast, a single two-level system always evolves to a mixed state with some excitation probability, regardless of the spectral form. Thus, our entanglement-based approach not only confirms the presence of the circular Unruh effect but also reveals spectral features that a single detector method cannot discern.

The steady-state entanglement phenomenon is significantly different from other phenomena related to the dynamics of quantum entanglement \cite{Valentini1991,Reznik2003,Steeg2009,Doukas2010,Salton-Man2015,martin2015,martin2018,Henderson2018,Zhjl2020,Zhjl2021,Benatti2004, JHu2015}, which could, in principle, be employed to detect the Unruh effect. 
However, these dynamical processes occur on extremely short timescales and depend on the initial state. In contrast, the entanglement induced by centripetal acceleration found here is long lived and initial-state independent, thereby enhancing the feasibility of experimental observation. Furthermore, phenomena such as entanglement harvesting \cite{Reznik2003, Salton-Man2015, Doukas2010, Zhjl2020, Zhjl2021}, entanglement dynamics \cite{Benatti2004,JHu2015}, the Lamb shift \cite{Audretsch95, Arya2023}, and Berry's phase \cite{Martin11, Hu12, Arya2022}, which could be utilized to detect the Unruh effect, already exist for inertial quantum systems, with the Unruh effect providing only a correction. In contrast, the steady-state entanglement unveiled here is a novel effect exclusively precipitated by the circular Unruh effect. We also show in detail that molecules are suitable candidates for the neutral but polarizable quantum systems considered here.

 At this point, it is important to note that achieving controllable generation of quantum entanglement with molecules has been extremely challenging and has only recently been realized following significant breakthroughs \cite{molecule1, molecule2}. By attaching a two-molecule system to an optically levitated hyperfast rotating nanoparticle, the steady-state entanglement observed here can be promisingly detected.
Hereafter, units with $\hbar=c=\epsilon_0=k_B=1$ are used, where $\hbar$ is the reduced Planck constant, $c$ is the speed of light, $\epsilon_0$ is the vacuum permittivity, and $k_B$ is the Boltzmann constant.

\emph{Physical model}. We consider a quantum system consisting of two neutral but polarizable two-level subsystems coupled with fluctuating electromagnetic fields in vacuum. As shown in Fig.~\ref{model}, the two subsystems rotate synchronously at the ends of a diameter of a circular orbit with  radius $R$ and  angular frequency $\Omega$. We work in the cylindrical coordinate system, where the trajectories of the two subsystems can be described as
\begin{eqnarray}\label{trajectories}
&&r_1(t)=R,\;\;\;\;\;\;\theta_1(t)=\Omega t,\;\;\;\;\;\qquad z_1(t)=0,\nonumber\\
&&r_2(t)=R,\;\;\;\;\;\;\theta_2(t)=\Omega t+\pi,\;\;\;\;\,~z_2(t)=0,
\end{eqnarray}
where $t$ is the coordinate time. 
The ground and excited states are labeled as $|0\rangle$ and $|1\rangle$, respectively, and the 
transition frequency between the two levels is $\omega_0$ in the proper frame of the two-level subsystem. Electric dipole transitions between the two levels $|0\rangle$ and $|1\rangle$ are allowed. For simplicity, we assume that the two-level subsystems are polarizable in the direction parallel to the axis of rotation. 
In the laboratory frame,  the interaction Hamiltonian between the two-level subsystems and the vacuum electromagnetic fields takes the form
\bea\label{int-Hamiltonian}
H_I=-\sum_{\alpha=1}^{2} D_{z}^{(\alpha)}[E_{z}(t,{\bm  x}_{\alpha})-R \Omega B_{r}(t,{\bm  x}_{\alpha})]\,,
\eea
where 
$\alpha=1,2$ labels the two subsystems, $D_i$ is the electric dipole moment, ${E_i}$ is the electric field strength, and $B_{r}$ is the magnetic induction strength. 
See Sec.~I of the supplemental material \cite{Supplemental} for more details on the derivation of the interaction Hamiltonian in the laboratory frame.

In the laboratory frame, the dynamics of the total system $\rho_{\rm tot}(t)$, consisting of the two two-level subsystems and the fluctuating electromagnetic fields,  satisfies the Liouville-von Neumann equation,
\bea\label{LvN-equation}
\frac{d}{d t}\rho_{\rm tot}(t)=-i[H_I(t),\rho_{\rm tot}(t)].
\eea
The total density matrix $\rho_{\text{tot}}(t)$ can be expressed in integral form as
$\rho_{\text{tot}}(t)=\rho_{\text{tot}}(0)-i\int_{0}^{t}[H_I(s),\rho_{\text{tot}}(s)]ds$.
Substituting this expression into Eq. \eqref{LvN-equation} and taking trace over the quantum fields, we obtain
\bea\label{eq4}
\frac{d}{d t}\rho(t)=-\int_{0}^{t}\text{Tr}_F[H_I(t),[H_I(s),\rho_{\text{tot}}(s)]]ds,
\eea
where $\rho(t)$ is the reduced density matrix of the quantum system consisting of two two-level subsystems. Since the coupling is weak 
and the number of degrees of freedom of the fluctuating electromagnetic fields is large, the density matrix $\rho_{\text{tot}}$  can be approximated as  $\rho_{\text{tot}}(s)\approx\rho(s)\otimes|0\rangle_{F}{}_{F}\langle0|$, where $|0\rangle_{F}$ is the vacuum state of the quantum fields.  Note that this approximation does not imply that there are no excitations in the reservoir caused by the reduced quantum system but rather that the environmental excitations decay over a timescale that is not resolved \cite{Breuer}.
Plugging the tensor product of $\rho_{\rm tot}$ and the interaction Hamiltonian \eqref{int-Hamiltonian} into Eq.~\eqref{eq4}, we obtain an integro-differential equation describing the dynamics of the reduced system,
\bea\label{m}
\frac{d\rho(t)}{d  t} =\sum\limits_{\alpha,\varrho=1}^{2}\int_{0}^{t}d s\Big\{G_{+}^{(\alpha\varrho)}(t-s)\big[D_{z}^{(\varrho)}(s)\rho(s)D_{z}^{(\alpha)}(t)
\nn\\-D_{z}^{(\alpha)}(t)D_{z}^{(\varrho)}(s)\rho(s)\big]\Big\}+\text{H.c.},
~~~
\eea
where ``H.c." denotes Hermitian conjugation, and
\bea\label{field correlation function}
G_{+}^{(\alpha\varrho)}(t-s)~~~~~~~~~~~~~~~~~~~~~~~~~~~~~~~~~~~~~~~~~~~~~~~~~~~~\nn\\
={}_F\big\langle0\big| \left[E_{z}\big(t,{\bm x}_{\alpha}(t)\big)-R\Omega B_{r}\big(t,{\bm x}_{\alpha}(t)\big)\right]~~~~~\nn\\
\times\left[E_{z}\big(s,{\bm x}_{\varrho}(s)\big)-R\Omega B_{r}\big(s,{\bm x}_{\varrho}(s)\big)\right]\big|0\big\rangle_F ~~
\eea
are the two-point functions of the electromagnetic fields in vacuum. Here, the two-point functions are functions of $t-s$, so they are invariant under temporal translations.

\emph{Steady-state entanglement}. Now, we investigate the entanglement of the rotating quantum system, specifically focusing on the possibility of steady-state entanglement. By solving the integro-differential equation \eqref{m} through a Laplace transform and employing the final value theorem, we obtain the steady state. See Sec.~II of the supplemental material \cite{Supplemental} for the derivation and the explicit form of the steady-state density matrix. It is important to note that the Markov approximation is not applied in this calculation.

We measure the entanglement of the quantum system using concurrence \cite{W. K. Wootters}, which is 0 when the quantum system is separable and 1 when it is maximally entangled. 
In Fig.~\ref{C(inf)(Om,R)}, we show the numerical result of the steady-state concurrence as a function of the rotation angular frequency and the orbital radius. 
When the rotation angular frequency is sufficiently large while the orbital radius is sufficiently small, i.e.,  $\Omega>\omega_{0}$ and $R\omega_{0}\ll1$, the quantum system can achieve  steady-state entanglement, which is independent of the initial state. Moreover, the leading term of the steady-state concurrence when $R\omega_{0}\ll1$ is calculated to be 
\bea\label{C0}
C\approx1\bigg/{\left[1+\frac{2 \Omega  \left(5 \omega_0^4+5 \omega_0^2 \Omega ^2+2 \Omega ^4\right)}{(\Omega -\omega_0)^3 \left(2 \omega_0^2+\omega_0\Omega +2 \Omega ^2\right)}\right]}.\hspace{0.6cm}
\eea
See Sec.~III of the supplemental material \cite{Supplemental} for the derivation of Eq. \eqref{C0}. We emphasize that the steady-state entanglement is independent of the initial state of the quantum system. This implies that it is purely induced by the centripetal acceleration and can thus be regarded as evidence of the circular version of the Unruh effect.
\begin{figure}[!htbp]
\centering
\includegraphics[width=0.4\textwidth]{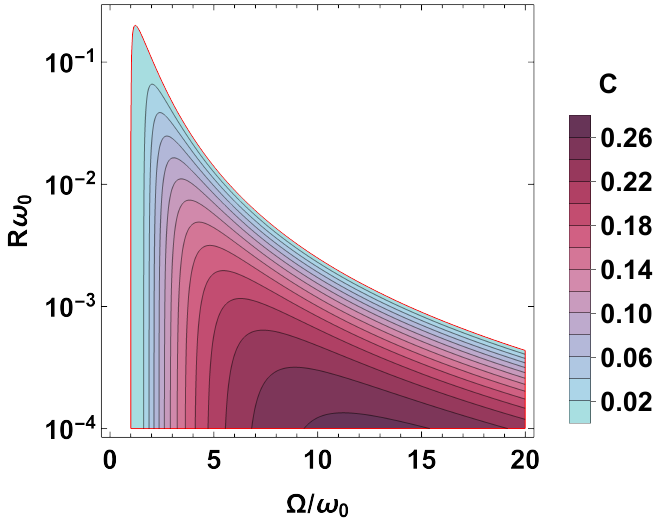}
\caption{\label{C(inf)(Om,R)} The steady-state concurrence as a function of $\Omega/\omega_{0}$ and $R\omega_{0}$. Note that the concurrence is fully determined by $\Omega/\omega_{0}$ and $R\omega_{0}$ with no dependence on other physical parameters.}
\end{figure}

\emph{Relaxation time}. Besides the degree of entanglement, another important factor for experimental detection is  the timescale for the quantum system to reach the steady state, i.e., the relaxation time.  When $\Omega>\omega_{0}$ and $R\omega_{0}\ll1$, i.e., when the steady-state entanglement exists, an approximate expression of the relaxation time can be obtained by analyzing the poles of the Laplace-transformed density matrix $\widetilde{\rho}(z)$ as
\bea\label{relaxation time1}
t_{\rm re}\sim\frac{1}{|d|^2R^2 \Omega^5},
\eea
where $d=\langle1|D_{z}|0\rangle$ is the element of the transition dipole moment matrix. See Sec.~IV of the supplemental material~\cite{Supplemental} for the derivation of the approximated relaxation time.

\emph{Experimental feasibility}. In the following, we propose detecting the steady-state entanglement induced by centripetal acceleration with the state-of-the-art technologies in optomechanics. Here, let us note that the following three requirements must be fulfilled: 
(1) a rotation angular frequency larger than the transition frequency of the two-level system, i.e., $\Omega/\omega_0>1$; (2) an orbital radius much smaller than the transition wavelength, i.e., $R\omega_0\ll1$; and 
(3) a reasonable relaxation time $t_{\rm re}$, i.e., the time required for the quantum system to attain steady-state entanglement. 
The first two conditions ensure the existence of steady-state entanglement, while the third condition ensures that the detection is feasible.

First, the rotation angular frequency $\Omega$ should be larger than the transition frequency $\omega_0$. Recently, hyperfast rotation of optically levitated nanoparticles has been achieved by transferring the spin angular momentum of light to the mechanical angular momentum of the particle \cite{Reimann2018,Li2018,Zhang2021}. To the best of our knowledge, the highest rotation frequency achieved with this mechanism so far is $\Omega\approx 6$ GHz for nanoparticles with an average radius $R=95$ nm \cite{Zhang2021}. 
These works suggest that one may attach the two-level systems to an optically driven hyperfast rotating nanoparticle to acquire a large angular frequency, and the question now is whether the transition frequency from an electric dipole allowed transition can be smaller than the current record rotation angular frequency. 
Natural quantum systems, e.g., atoms and molecules, are multilevel systems. However,  in vacuum, they can be effectively treated as two-level systems with only the ground state and the first excited state, as transitions to higher energy levels rarely occur.
For atoms, the energy level spacing between the ground state and the first excited state is typically on the order of $\sim 1$~eV, which corresponds to a transition frequency in the optical frequency regime ($\sim 10^{15}$~Hz). Obviously,  it is much larger than the currently achievable  rotation angular frequency. However, for molecules, the typical value of the transition frequency between rotational states is in the microwave regime, i.e., $\omega_0 \sim 10^{9}-10^{13}$~Hz~\cite{mol-phys}. 
This indicates that the condition $\Omega/\omega_0>1$ can be achieved by choosing molecules with appropriate transition frequencies. 
Additionally, artificial two-level atoms, such as superconducting qubits, have transition frequencies typically in the GHz regime. However, the typical size of superconducting qubits falls within the micrometer range, making them unsuitable for fabrication on nanoparticles to achieve hyperfast rotation angular frequencies. Therefore, for the reasons outlined above, we will assume the two-level systems to be molecules in the following discussion.

Second, the orbital radius $R$ should be much smaller than the transition wavelength $\omega_0^{-1}$.  Simultaneously, it must be larger than the size of the two-level systems, as they are supposed to rotate at the ends of a diameter of a circular orbit. Typically, for diatomic molecules, the size characterized by the bond length is on the order of $\sim 10^{-10}$ m, while the transition wavelength of the ground state and the first excited state is on the order of $\sim 10^{-1}$~m. 
Thus, the orbital radius can be much smaller than the transition wavelength and still be significantly larger than the size of the molecules.

Third, the relaxation time should be reasonable for experimental feasibility, ideally within a few hours ($t_{\rm re}\sim10^4$~s) or less.  
According to the approximated expression Eq.~\eqref{relaxation time1}, the relaxation time can be decreased by increasing the rotation angular frequency $\Omega$ and the orbital radius $R$. However,  for an optically driven nanoparticle considered here, $\Omega$ and $R$ are not independent. For such a nanoparticle, stable rotation is achieved when the driving torque
 $M_0$ equals to the drag torque $M_d$ \cite{Reimann2018,Zhang2021}. The driving torque  $M_0$ is proportional to the intensity of the trapping light ${\cal I}_0$ and the radius of the nanoparticle  $R$, i.e., $M_0\propto {\cal I}_0 R$.  The drag torque $M_d$ is proportional to the rotation angular frequency $\Omega$, the moment of inertia of the particle $I\propto R^2$, and the damping rate of the rotation motion $\gamma_d \propto R^2$ \cite{Fremerey1982}, i.e., $M_d\propto \gamma_d I \Omega$.
Therefore, the rotation angular frequency of the nanoparticle $\Omega\propto \frac{{\cal I}_0 R}{\gamma_d I}\propto \frac{{\cal I}_0 }{R^3}$, 
and the relaxation time
\bea\label{}
t_{\rm re}\propto\frac{1}{R^2 \Omega^5}\propto R^{13}.
\eea
Thus, for a quantum system consisting of two two-level subsystems attached to a hyperfast optically rotating nanoparticle, an effective way to reduce the relaxation time is to decrease the size of the nanoparticle. 
For the experimental setup in Ref. \cite{Zhang2021},  the rotation angular frequency is expected to reach $\sim10^{4}$~GHz if the radius of the nanoparticle is reduced to $4.75$~nm, and the relaxation time is then estimated to be on the order of $\sim 10^{4}$~s (a few hours).

As discussed above, the entanglement induced by centripetal acceleration can hopefully be detected if we attach two molecules to an optically levitated hyperfast rotating nanoparticle. 
Prior to embarking on experimental detection efforts, it is crucial to address several pertinent issues. Our study initially posits an ideal vacuum environment, a condition that simplifies theoretical modeling but diverges from practical experimental contexts where thermal fluctuations and dephasing are inevitable. Despite these challenges, our analysis  (see Sec.~V of the supplemental material \cite{Supplemental} for details) demonstrates that when the environment temperature is sufficiently low and the parameter characterizing dephasing is sufficiently small the impact of thermal fluctuations and the dephasing channel on concurrence becomes negligible. Furthermore, these conditions do not significantly alter the order of magnitude of the relaxation time. This suggests that the entanglement's robustness in such a quantum system under centripetal acceleration can be preserved in the face of environmental perturbations, provided the specific conditions regarding temperature and dephasing are met.  For an explicit example,  let us consider a scenario where the rotation angular frequency of the quantum system ($\Omega$) is $\sim10^4$ GHz, a value within the realm of feasibility. 
 Additionally, 
if the transition frequency ($\omega_0$) is  $\sim10^3$ GHz, and provided that the environmental temperature is maintained below 10 K with the dephasing transition coefficient being under  $10^{-2}\;\text{s}^{-1}$,  the adverse effects of thermal fluctuations and dephasing on the steady-state concurrence  are significantly mitigated. Specifically, their negative impact on concurrence is less than $5\%$ compared to the situation where the environment is an ideal vacuum and no dephasing channel exists. This scenario underscores the potential for achieving and maintaining significant levels of entanglement in quantum systems under specific, controlled conditions, even in the presence of environmental perturbations.  The second 
is that the energy levels of the molecules must be protected when they are attached to the particle. As proposed in Ref.~\cite{Lochan20}, this can be realized, e.g., by enclosing the molecules in a C$_{60}$ fullerene~\cite{c60}. The C$_{60}$ fullerene containing the molecule can be fabricated on the nanoparticle by, e.g., the wet chemical method \cite{Boul99}, and the ion bombardment method~\cite{Karin2019}. The third is how to measure the entanglement of the two-molecule system. In fact, the most straightforward way is a complete tomographic reconstruction of the quantum state of the system. See Ref.~\cite{tomography} for a review. Recently, the entanglement between the rotational states of two individual molecules has  been generated, and the entanglement is certified by reconstructing some of the density matrix elements \cite{molecule1,molecule2}. These works indicate that the entanglement between two rotating molecules found here may be similarly detected. 

Finally, another promising platform for detecting the steady-state entanglement induced by centripetal acceleration is circularly moving optical tweezers. Recently, fully loaded two-dimensional arrays of individual atoms with arbitrary geometries have been achieved using a real-time control system and moving optical tweezers \cite{Barredo2016}, where the speed of the moving optical tweezers can reach approximately  $10~{\rm nm/\mu s}$. Given that the radius of the orbit considered here is on the order of  $10~{\rm nm}$,
 it is expected that the angular velocity can reach $10^6~{\rm Hz}$, which is still 6-7 orders of magnitude smaller than the rotational angular velocity required in our proposal. Nevertheless, this platform is promising since the measurements and state tomography are much easier to perform.

\emph{Summary}. We have found that a quantum system consisting of two neutral polarizable two-level subsystems under centripetal acceleration in vacuum can acquire steady-state entanglement, when the rotation angular frequency is larger than the transition frequency and the orbital radius is much smaller than the transition wavelength. The steady-state entanglement is independent of the initial state, so it is induced by the centripetal acceleration, and can be regarded as evidence of the circular version of the Unruh effect. Detailed analysis shows that the phenomenon can promisingly be detected by attaching two molecules to a hyperfast rotating optically levitated nanoparticle with state-of-the-art optomechanical technologies.

\begin{acknowledgments}

We extend our gratitude to the anonymous referee for bringing the moving optical tweezers to our attention, to Wenting Zhou for helpful discussions, and to Wenjie Wang for assistance with graphics. This work was supported in part by the NSFC under Grants No. 11805063, No. 12075084, and No. 12375047; the Innovative Research Group of Hunan Province under Grant No. 2024JJ1006; the Hunan Provincial Natural Science Foundation of China under Grants No. 2020JJ3026 and No. 2023JJ40515; and the Scientific Research Program of Education Department of Hunan Province of China under Grant No. 22B0762.
\end{acknowledgments}

\end{document}